\documentclass[aps,prl,preprint,superscriptaddress]{revtex4-1}
\usepackage{upgreek}
\usepackage{amssymb, amsmath}
\usepackage{bm}
\usepackage{pifont}
\usepackage{graphicx}
\usepackage[usenames]{color}
\usepackage{makecell}% make cells in tables
\usepackage[T1]{fontenc}%for some references

\begin{document}

\title{Generalization of the Nested Wilson Loop Formalism in Topological Dirac Semimetals with Higher-order Fermi Arcs}

\author{Hui Zeng}
\affiliation{State Key Laboratory of Low-Dimensional Quantum Physics, Department of Physics, Tsinghua University, Beijing 100084, China}

\author{Wenhui Duan}
\affiliation{State Key Laboratory of Low-Dimensional Quantum Physics, Department of Physics, Tsinghua University, Beijing 100084, China}
\affiliation{Institute for Advanced Study, Tsinghua University, Beijing 100084, China}
\affiliation{Frontier Science Center for Quantum Information, Beijing 100084, China}
\affiliation{Collaborative Innovation Center of Quantum Matter, Beijing 100871, China}

\author{Huaqing Huang}
\email[Corresponding author: ]{huaqing.huang@pku.edu.cn}
\affiliation{School of Physics, Peking University, Beijing 100871, China}
\affiliation{Collaborative Innovation Center of Quantum Matter, Beijing 100871, China}
\affiliation{Center for High Energy Physics, Peking University, Beijing 100871, China}

\date{\today}

\begin{abstract}
 We generalize the nested Wilson loop formalism, which has been playing an important role in the study of topological quadrupole insulators, to two-dimensional (2D) and 3D nonsymmorphic materials with higher-order topology. In particular, certain 3D Dirac semimetals exhibit 1D higher-order Fermi arc (HOFA) states localizing on hinges where two surfaces meet and connecting the projection of the bulk Dirac points. We discover that the generalized nested Berry phase (gNBP) derived from this formalism is the bulk topological indicator determining the existence/absence of HOFAs, revealing a direct bulk-hinge correspondence in 3D Dirac semimetals. Finally, we study the Dirac semimetals NaCuSe and KMgBi based on first-principles calculations and explicitly show that the change in gNBP adjacent to the Dirac point corresponds to the termination of HOFAs at the projection of Dirac points on the hinge. Our findings not only improve the understanding of the bulk-hinge correspondence in topological Dirac semimetals but also provide a general formalism for studying the higher-order topology in nonsymmorphic systems.
\end{abstract}

%\keywords{}

\maketitle

\textit{Introduction.}---The bulk-boundary correspondence, which is the novel feature of topological states of matter, relates the nontrivial bulk topology to the presence of gapless states at boundaries \cite{RevModPhys.82.3045, RevModPhys.83.1057}. Recently, the notion of bulk-boundary correspondence has been extended to higher-order forms where topologically protected gapless modes emerge at even lower-dimensional boundaries, such as corners or hinges of crystals \cite{bernevig2017quantizedNBP, bernevig2017detailedNBP, SongZhiDa2017R_symm_protected_HOTI, Fang2019Rotation_anomaly_TCI,  Guo2022NBP_in_Ta2M3Te5, khalaf2021BOTP, Ahn2019SW_review, Ahn2018SW_in_nodal_line, Park2019HOTI_in_TBG, Pan2022SW_in_Xene, Xie2021Review_band_HOTI, PhysRevB.98.201114}. In seminal works \cite{bernevig2017quantizedNBP}, Benalcazar \textit{et al.} have introduced the topological quadrupole insulator (TQI) which features a quantized bulk quadrupole moment and zero-energy corner modes. Based on the nested Wilson loop approach \cite{bernevig2017quantizedNBP}, the quadrupole moment in a TQI is determined from the nested Berry phase (NBP), a topological indicator \cite{khalaf2021BOTP} associated with the bands formed by Wannier centers of electrons, which is in analogy to the conventional Berry phase of electronic bands in crystals \cite{PhysRevLett.62.2747,PhysRevB.47.1651}.

Recently, certain Dirac semimetals have been shown to exhibit unique higher-order bulk-hinge correspondence, i.e., 1D higher-order Fermi arc states localize on hinges where two surfaces meet and terminate at the projection of bulk Dirac points \cite{bernevig2020HOFA,SXL2022TopologicalHM,Nie2022KagomeHOFA, Lin2017Topo_Quad_SM, Wang2022HODSM_in_phononic, Xia2022Experimental_HODSM_in_phononic}. The HOFA was regarded as a direct consequence of the bulk topology which is characterized by the change in NBP of 2D slices adjacent to the Dirac point in the 3D Brillouin zone (BZ) of Dirac semimetals \cite{bernevig2020HOFA}. However, this was limited to the specific cases with fourfold rotational symmetry, which raises the question of whether the topological analysis based on NBPs of BZ slices is still valid for other Dirac semimetals with different symmetries. Moreover, it has been found that some nonsymmorphic crystals exhibit higher-order corner states but only have a specially defined NBP \cite{liu2022nonsymmorphic_photonic, Zhang2018nonsymmorphic_Sonic_NC, lin2020nonsymmorphic_Sonic_PRB}, thus posing an open question of how to identify the bulk topology to furnish the bulk-hinge correspondence for these Dirac semimetals with nonsymmorphic symmetries.

In this work, we generalize the nested Wilson loop formalism to nonsymmorphic systems with higher-order bulk-boundary correspondence. The generalized NBP (gNBP) derived from this formalism at 2D BZ slices indexed by $k_z$ is proved to be a bulk topological indicator that directly determines the existence/absence of HOFAs on hinges of Dirac semimetals with nonsymmorphic symmetries. Based on first-principles calculations, we further study two Dirac semimetals, NaCuSe and KMgBi, and explicitly demonstrate that the change in gNBP across the Dirac points is associated with the termination of HOFAs at the projection of Dirac points on the hinge.

\textit{Generalization of nested Wilson loop.}---We first introduce the generalization of the nested Wilson loops and NBP. For a 2D BZ in the $k_x$-$k_y$ plane or any 2D slices with fixed $k_z$ in a 3D BZ, the Wilson loop along the $k_x$ direction with the base point $\mathbf{k}$ is defined as,
\begin{eqnarray}\label{wilson}
\mathcal{W}_{x,\mathbf{k}}=\mathcal{P}\mathrm{exp}\left(-i\int_{C_\mathbf{k}}
\mathcal{A}_{x,\mathbf{k}}dk_x\right),
\end{eqnarray}
{where} $C_{\mathbf{k}}$ is a contour along $k_x$ from $\mathbf{k}$ to $\mathbf{k} + (2\pi, 0)$, $[\mathcal{A}_{x,\mathbf{k}}]^{mn}=-i\langle u_k^m|\partial_{k_x}|u_k^n\rangle$ is the Berry connection of occupied bands, and $\mathcal{P}$ is the path-ordering operator. Diagonalizing these Wilson loop operators, one arrives at
\begin{equation}\label{wilsoneig}
  \mathcal{W}_{x,\mathbf{k}}|\nu_{x,\mathbf{k}}^j\rangle=\exp[i2\pi \nu_{x}^j(k_y)]|\nu_{x,\mathbf{k}}^j\rangle,
\end{equation}
where eigenstates $|\nu_{x,\mathbf{k}}^j\rangle$ ($j=1,2,\cdots, N_{occ}$) have components $[\nu_{x,\mathbf{k}}^j]^n$, ($n=1,2,\cdots,N_{occ}$). The eigenphases $\nu_{x}^j(k_y)$ which is only a function of $k_y$, form Wannier bands over the 1D BZ $k_y\in(-\pi,\pi]$ and obey the identification $\{\nu_{x}^j(k_y)\}\equiv\{\nu_{x}^j(k_y)\mod 1\}$.

Due to certain symmetry constraints, such as mirror reflections, these Wannier bands can be naturally divided into two groups, i.e., two Wannier sectors, which are separated from each other by Wannier gaps in the Wannier band spectrum \cite{bernevig2017quantizedNBP}. The key insight behind the nested Wilson loop formalism is that, being gapped, the Wannier sectors can carry their own topological invariants. Here we generalize this idea by dividing the occupied energy bands into $N_s$ $(\geq2)$ subspaces according to the generic splitting of Wannier sectors constrained by symmorphic and/or nonsymmorphic symmetries,
\begin{equation}\label{w_s}
|w_{x, \mathbf{k}}^{s,j}\rangle=\sum_{n=1}^{N_{occ}}|u_{k}^{n}\rangle[\nu_{x, \mathbf{k}}^{s,j}]^{n}, \quad s=1,2,\cdots, N_s,
\end{equation}
where the superscript $s$ denotes the Wannier sectors, with $j\in \{1,2,\cdots, N_W=N_{occ}/N_s\}$ labeling the bands within a sector. %$N_W=N_{occ}/N_s$ is the number of Wannier bands in a sector.
Here we assume that $N_{occ}$ is an integer multiple of $N_s$, otherwise the Wannier spectrum is always gapless \cite{bernevig2017detailedNBP}.
The nontrivial nature of each Wannier sector is determined by the generalized Wannier-sector polarization (WSP),
\begin{equation}\label{eq: generalized NBP defination}
\begin{aligned}
&p_{y}^{\nu_x^{s}}=-\frac{1}{(2 \pi)^2} \int_{B Z} {\text{Tr}[\tilde{\mathcal{A}}_{y, \mathbf{k}}^{\nu_x^{s}}]} d^2 \mathbf{k}, \\
&[\tilde{\mathcal{A}}_{y, \mathbf{k}}^{\nu_x^{s}}]^{j,j'}=-i\langle w_{x, \mathbf{k}}^{s,j}|\partial_{k_{y}}| w_{x, \mathbf{k}}^{s,j'}\rangle.  \\
\end{aligned}
\end{equation}
Similar procedures of Eqs.~(\ref{wilson})-(\ref{eq: generalized NBP defination}) are also applicable to the $k_y$ direction, which bring up the Wilson loop operator $\mathcal{W}_{y,\mathbf{k}}$, Wannier bands $\nu_{y}^j(k_x)$, and generalized WSP $p_x^{\nu_y^s}$ along the $x$ direction. In practice, we numerically determine the Wilson loop operators by discretizing the BZ and the detailed procedure is thoroughly explained in Supplementary Material (SM) \footnote{\label{fn}See Supplementary Material for more details about the derivation of the generalized Wilson loop, symmetry constraints, and the numerical calculation, which include Refs.~\cite{VASP,wannier90,HopTB,irvasp,openmx}}. Under symmetry constraints, these generalized WSPs would take quantized values, which are expected to indicate the band topology, as discussed later. For convention, we choose the $i$-th WSP as an indicator so that the gNBP $\theta^{(s)}=2\pi (p_{x}^{\nu_y^{i}}, p_{y}^{\nu_x^{i}})$ are zeros in the trivial atomic limit.

\textit{Symmetry constraints and quantization.}---It is noted that Ref.~\cite{bernevig2017quantizedNBP} is the symmorphic case for 2D lattices with reflection symmetries $\mathcal{M}_x: x\rightarrow-x$ and $\mathcal{M}_y: y\rightarrow-y$, which enforce Wannier bands to occur in pairs
$\{\nu_\alpha^j(k_\beta)\}\overset{\mathcal{M}_{\alpha}}{=}\{-\nu_\alpha^j(k_\beta)\},
\{\nu_\alpha^j(k_\beta)\}\overset{\mathcal{M}_{\beta}}{=}\{\nu_\alpha^j(-k_\beta)\}$
with $\alpha,\beta\in\{x,y\}$ and $\alpha\neq\beta$. From this, it follows a natural splitting of occupied energy bands into $N_s=2$ subspaces based on two separable Wannier sectors $\nu_\alpha^{-}\in[-1/2,0)$ and $\nu_\alpha^{+}\in[0,1/2)$. The polarizations of two Wannier sectors obey $p_{\beta}^{\nu_\alpha^{+}}\overset{\mathcal{M}_{\alpha}}{=}p_{\beta}^{\nu_\alpha^{-}}, p_{\beta}^{\nu_\alpha^{\pm}}\overset{\mathcal{M}_{\beta}}{=}-p_{\beta}^{\nu_\alpha^{\pm}}$. Therefore, although the full space of occupied energy bands has trivial polarizations $p_\beta=p_{\beta}^{\nu_\alpha^{+}}+p_{\beta}^{\nu_\alpha^{-}}=0$, the WSPs take quantized values $p_{\beta}^{\nu_\alpha^{\pm}}\overset{\mathcal{M}_{\beta}}{=}0$ or $1/2 \mod 1$ for trivial or nontrivial cases, respectively. %Consequently,
In particular, nontrivial WSP $(p_{x}^{\nu_y^{-}},p_{y}^{\nu_x^{-}})=(1/2,1/2)$ implies the quantized edge polarization, fractional corner charge, and nontrivial quadrupole invariant $q_{xy}=2p_{x}^{\nu_y^{-}}p_{y}^{\nu_x^{-}}=1/2$ in topological quadrupole insulators \cite{bernevig2017quantizedNBP}.

The generalization of NBP is demanded in the cases of nonsymmorphic symmetry, where conventional NBP $\theta^{(2)}_{\beta}$ vanishes, but the existence of higher-order topological features indicates the urgent need for an essential topological indicator. In general, a nonsymmorphic symmetry operator of a layer group can be expressed as $\{\mathcal{O}_\alpha|\delta_\iota\}$, where $\mathcal{O}_\alpha$ is a two-fold rotation $\mathcal{C}_{2\alpha}$ or a reflection $\mathcal{M}_\alpha$ with $\alpha=x$ or $y$, and $\delta_\iota$ takes fractional values (in unit-cell units): $\delta_x=(\frac{1}{2},0,0), \delta_y=(0,\frac{1}{2},0)$, and $\delta=(\frac{1}{2},\frac{1}{2},0)$. For example, under glide reflections $\mathcal{G}_{\alpha}= \{\mathcal{M}_\alpha|\delta\}$, the Wannier bands pair up differently,
\begin{equation}\label{glide}
\begin{aligned}
    \left\{\nu_{x}^j(k_y) \right\} & \overset{\mathcal{G}_{x}}{=} \left\{-\nu_{x}^j(k_y) +1/2 \right\}, \\
    \left\{\nu_{x}^j(k_y) \right\} & \overset{\mathcal{G}_{y}}{=} \left\{\nu_{x}^j(-k_y) +1/2 \right\}.
\end{aligned}
\end{equation}
Hence, due to the constraints of $\mathcal{G}_{x}$ and $\mathcal{G}_{y}$, the Wannier bands always occur in quadruple: $\nu_x(k_y)$, $1/2-\nu_x(k_y)$, $-\nu_x(-k_y)$, $1/2+\nu_x(-k_y)$. This provides a natural splitting of Wannier bands into four sectors (i.e., $N_s=4$): $\nu_x^s(k_y)\in [-\frac{1}{2}+\frac{s-1}{4}, -\frac{1}{2}+\frac{s}{4})$ with $s=1,2,3,4$ when $N_{occ}=4n$. Consequently, the quartern WSPs obey
\begin{equation}\label{glide polar}
\begin{aligned}
    p^{\nu_x^1}_y \overset{\mathcal{G}_{x}}{=} p^{\nu_x^2}_y+\frac{n}{2} \overset{\mathcal{G}_{y}}{=} -p^{\nu_x^4}_y \overset{\mathcal{G}_{x}}{=} -p^{\nu_x^3}_y +\frac{n}{2} &\text{ mod}(1).
\end{aligned}
\end{equation}
Although the constraints from $\mathcal{G}_{x}$ and $\mathcal{G}_{y}$ alone are not enough to quantize $p^{\nu_x^s}_y$ itself, a composite WSP of two quarter Wannier sectors take quantized values which depends only on the parity of $n$: $p^{\nu_x^1}_y+p^{\nu_x^3}_y=p^{\nu_x^2}_y+p^{\nu_x^4}_y=\frac{1}{2}n \mod 1$.
This has been used to characterize the so-called ``anomalous'' quadrupole insulators \cite{Zhang2018nonsymmorphic_Sonic_NC,lin2020nonsymmorphic_Sonic_PRB}, which cannot undergo a topological phase transition to trivial insulators unless the fundamental nonsymmorphic symmetries are broken.

More importantly, we found that under some combination of nonsymmorphic and/or symmorphic symmetries, the generalized WSP would take quantized values, which signals the higher-order topology when the conventional WSP vanishes. For example, the $p4/nmm$ layer group includes both reflection $\mathcal{M}_{\alpha}$ with $\alpha\in\{x,y\}$ and off-centered rotation $\{\mathcal{C}_{2\alpha}|\delta\}=\mathcal{G}_{\beta}\cdot \mathcal{M}_z$ ($\beta \neq \alpha)$ \cite{PhysRevB.95.075135}. Because the out-of-plane reflection symmetry $\mathcal{M}_z$ would not affect the Wilson loop along in-plane directions, the constraints generated by $\{\mathcal{C}_{2\alpha}|\delta\}$ are the same as those generated by glide reflection $\mathcal{G}_{\beta}$ [see Eqs. (\ref{glide}) and (\ref{glide polar})], indicating a quadrisection of Wannier bands. $\mathcal{M}_{\alpha}$ further quantizes the quartern WSP,
\begin{equation}\label{p4nmm}
p_{\alpha}^{\nu_{\beta}^s} = 0\; \mathrm{or}\; \frac{1}{2} \mod 1, \quad (s=1,2,3,4)
\end{equation}
which still satisfy the constraint of Eq.~(\ref{glide polar}).

\begin{table}[t]%[tbhp]
    \renewcommand \arraystretch{1.5}
    \centering
    \caption{Layer groups that admit higher-order topology with quantized gNBP as a topological indicator.} %Layer groups with higher-order topology characterized by quantized gNBP.}
    \label{table1}
    \begin{tabular}{c c}
         \hline
         \hline
         Quantized gNBP & Layer groups  \\
         \hline
         \makecell[c]{$\theta^{(2)}$ }  & \makecell[c]{$p2/m11, p222, pmm2, pmmm, p422,$\\ $p4mm, p$-$42m, p$-$4m2, p4/mmm$} \\
         \hline
         \makecell[c]{$\theta^{(4)}$ }  & \makecell[c]{$c2/m11, c222, cmm2, pban, pman,$ \\ $pmmn, cmmm, cmme, p4/nbm, p4/nmm,$ \\ $p$-$31m, p$-$3m1, p622, p6mm, p6/mmm$} \\
         \hline
         \hline
    \end{tabular}
\end{table}

It is worth mentioning that the generalized WSP applies not only to orthogonal systems with nonsymmorphic symmetries but also to systems with non-orthogonal primitive cells. 
For example, in $\mathcal{C}_3$- or $\mathcal{C}_6$-symmetric systems, one can redefine a non-primitive rectangular cell to restore the periodicities along two orthogonal directions. A glide symmetry emerges by combining reflection with fractional translation $\{\mathcal{E}| \frac{1}{2}, \frac{1}{2},0\}$ in units of the redefined cell \cite{PhysRevB.102.195202}. Then, the generalized WSP can be calculated analogically. We systemically considered all 80 layer groups, which describe the spatial symmetry of 2D spinless systems \cite{delaFlor2021LayerBilbao}. The results are summarized in Table~\ref{table1} (see SM for more details \footnotemark[\value{footnote}]).

\begin{figure}
  \centering
  \includegraphics[width=1.0\linewidth]{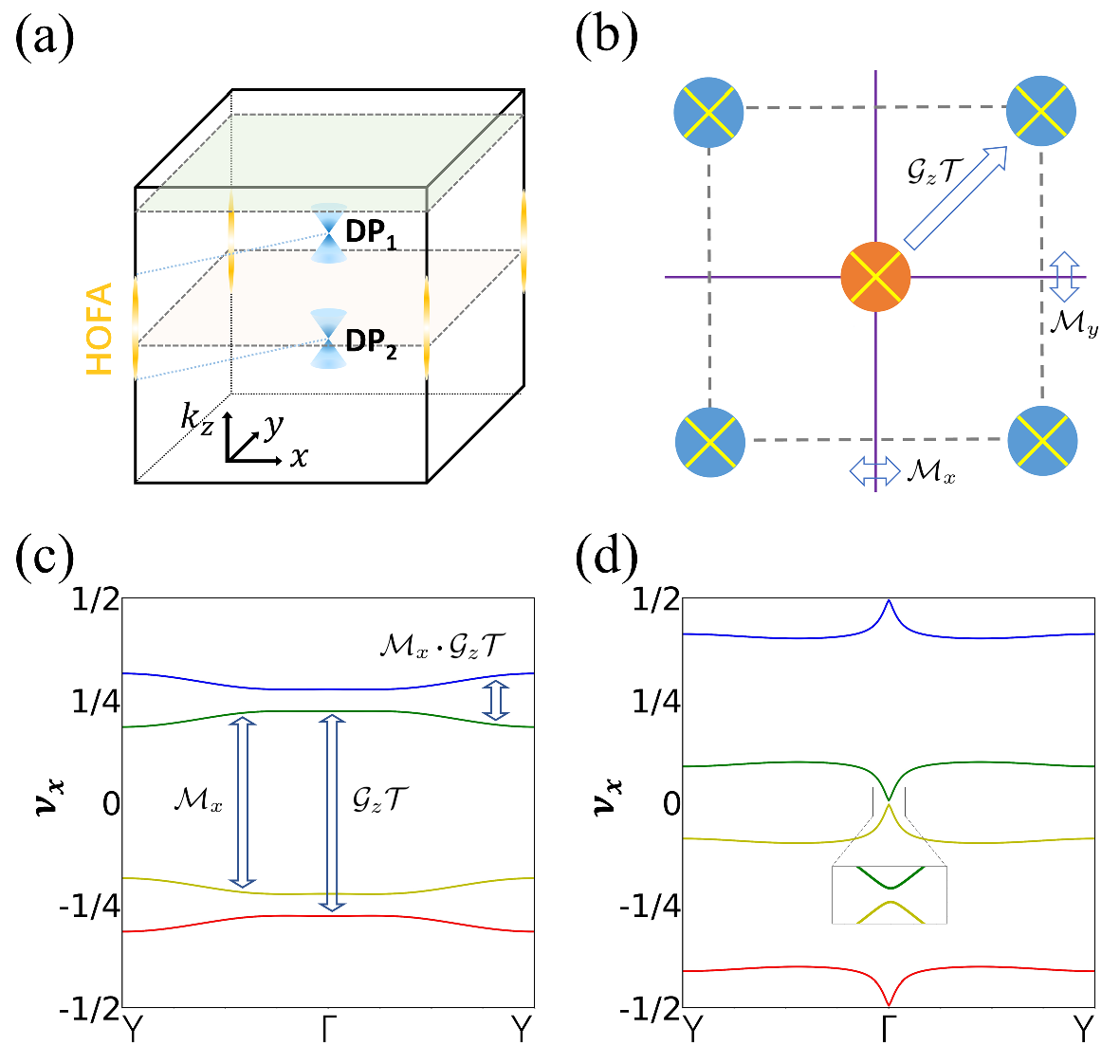}
  \caption{(a) Schematic illustration of a Dirac semimetal with a pair of Dirac points in space group $P4/nmm$. The HOFAs exist between the projections of two Dirac points on side hinges. The symmetries that keep generalized WSP invariant in slices at fixed $k_z\neq 0,\pi$ can be mapped to the layer group $p4/nmm$ by regarding $\mathcal{M}_z\mathcal{T}$ in the former as $\mathcal{M}_z$ in the latter. (b) Each $k_z$ slice is viewed as an effective 2D subsystem with the symmetry of layer group $p4/nmm$. The composite operator combing $\mathcal{G}_z= \{\mathcal{M}_z|\delta\}$ and $\mathcal{T}$ leads to a fractional transition $\{\mathcal{E}|\frac{1}{2}, \frac{1}{2},0\}$. (c),(d) Wannier bands of two $k_z$ slices at opposite sides of a Dirac point, where the quartern WSPs are 0 and 1/2, respectively. Four Wannier sectors are related by $\mathcal{M}_{x}$, $\mathcal{G}_z\mathcal{T}$, and $\mathcal{M}_{x}\cdot\mathcal{G}_z\mathcal{T}$ symmetries.}
  \label{fig1:model illustrate}
\end{figure}

In particular, the generalized WSP for layer groups can also extend to describe the higher-order topology of 2D insulating planes of 3D BZ indexed by $k_z\neq 0,\pi$ which are usually described by magnetic layer groups. In general, the time-reversal $\mathcal{T}$ and reflection $\mathcal{M}_z$ transfer $k_z$ to $-k_z$. Hence, symmetries that keep the generalized WSP invariant in generic slices ($k_z\neq 0$ or $\pi$) are in-plane symmetries $g_{xy}$ inherited from its 3D bulk and the composite symmetries combining $g_{xy}$, $\mathcal{T}$, and $\mathcal{M}_z$. In fact, a one-to-one mapping between the allowed symmetries of a $k_z$ slice and a layer group can be constructed by regarding $\mathcal{M}_z\mathcal{T}$ in the former to $\mathcal{M}_z$ in the latter. This is exemplified by considering a minimal model for the nonsymmorphic space group $P4/nmm$ that describes a 3D Dirac semimetal with a pair of Dirac points on the $k_z$ axis, as shown in Fig.~\ref{fig1:model illustrate}. Details of the tight-binding model are presented in SM \footnotemark[\value{footnote}]). The symmetry of a generic $k_z\neq 0,\pi$ slice are $\mathcal{M}_x$, $\mathcal{M}_y$, and $\mathcal{G}_z\mathcal{T}$, which give rise to the nonsymmorphic symmetry operator $\{\mathcal{C}_{2\beta}\mathcal{T}|\delta\} =\mathcal{M}_{\alpha}\cdot \mathcal{G}_{z} \mathcal{T}=\mathcal{G}_{\alpha}\cdot \mathcal{M}_{z} \mathcal{T}$. {Since $\mathcal{M}_z\mathcal{T}$ only reverses the Wannier band $\nu_\alpha(k_\beta) \rightarrow\nu_\alpha(-k_\beta)$ but does not affect Wannier gaps within a $k_z$ slice}, the $k_z$ slice exhibits a similar quadrisection of Wannier bands and quantization of generalized WSP as the layer group $p4/nmm$, which has been studied above. Specifically, the four Wannier sectors: $\nu_x(k_y)$, $1/2-\nu_x(-k_y)$, $-\nu_x(k_y)$, $1/2+\nu_x(-k_y)$ are related by $\mathcal{M}_x$, $\mathcal{G}_z\mathcal{T}$ and $\mathcal{M}_{x} \cdot\mathcal{G}_{z} \mathcal{T}$, as shown Fig.~\ref{fig1:model illustrate}(c,d). Although Wannier bands exhibit distinct behaviors for $k_z$-slices below or above a Dirac point, the conventional NBP $\theta^{(2)}$ vanishes for $k_z$-slices on both sides of the Dirac point, indicating its failure in diagnosing band topology.

Remarkably, with increasing $k_z$ across a Dirac point, the quartern WSP undergoes a sudden jump by 1/2, implying a topological phase transition for 2D $k_z$-slice subsystems. Meanwhile, we find that the Dirac semimetal exhibits higher-order Fermi arcs (HOFAs) terminating at the projection of the bulk Dirac points on side hinges. Such HOFAs can be viewed as topological bulk-hinge correspondence for Dirac semimetals \cite{bernevig2020HOFA}. Viewing each $k_z$ slice of the 3D BZ as an effective 2D subsystem, the Dirac point is equivalent to the critical point between 2D trivial and higher-order topological insulators with corner states. Therefore, our proposed generalized WSP serves as a topological indicator to determine the presence/absence of HOFAs. Our approach is also consistent with the previous classification of Dirac points in Refs.~\cite{ Fang2021FillingAnomalyC4,Fang2021ClassificationDSM} where the termination of HOFA is signaled by the change of filling anomaly for 2D $k_z$ slices on either side of the Dirac point \footnotemark[\value{footnote}].

\begin{figure}
  \centering
  \includegraphics[width=1.0\columnwidth]{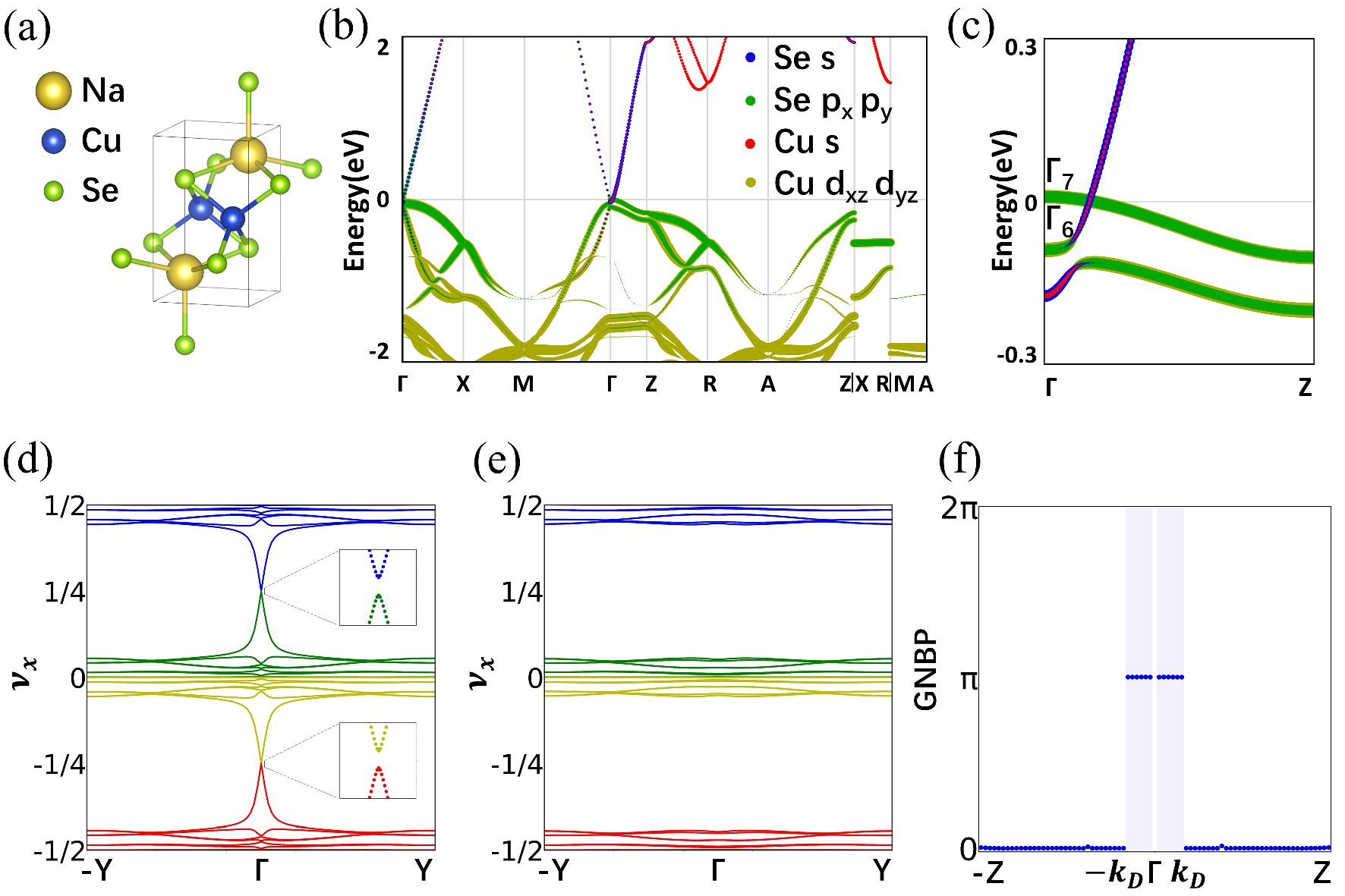}
  \caption{(a) Crystal structure of NaCuSe with $P4/nmm$ symmetry, where Na, Cu, and Se atoms occupy the $2c$, $2a$, and $2c'$ Wyckoff positions, respectively. (b) Orbital-resolved band structure of pristine NaCuSe, where the valence and conduction bands cross along $\Gamma$-$Z$ forming a Dirac point near the $\Gamma$ point. (c) Band structures along $\Gamma$-$Z$ of NaCuSe under a tensile strain of 6\% along the $z$-axis.  (d) and (e) are Wannier bands on $k_z = 0.04\pi$ and $k_z=0.20\pi$ slices which are on opposite sides of the Dirac point ($k_z=k_D$). The four Wannier sectors are marked in different colors. (f) The gNBP component $\theta^{(4)}_y$ of $k_z$ slices.}
  \label{fig:NaCuSe Structure}
\end{figure}

\textit{Material Realizations.}---As a concrete material example, we propose NaCuSe as a promising candidate for Dirac semimetal with nontrivial gNBP and HOFAs. As shown in Fig.~\ref{fig:NaCuSe Structure}(a), NaCuSe crystallines a tetragonal structure with space group $P4/nmm$ (No. 129), where Na, Cu, and Se atoms occupy the $2c$, $2a$, and $2c'$ Wyckoff positions, respectively. Our first-principles calculations show that around the Fermi energy, the electronic band structure of NaCuSe is mainly contributed by the Se $s$ and $p_{x,y}$, and Cu $d$ orbitals, as shown in Fig.~\ref{fig:NaCuSe Structure}(b). Because of the preserved $\mathcal{PT}$ symmetry, every band is doubly degenerate. It is noteworthy that the $s$-dominated band is lower than the Se-$p$ band at $\Gamma$, resulting in an inverted band structure. Since the two bands belong to $\Gamma_6$ and $\Gamma_7$ irreducible representations with respect to $\mathcal{C}_{4z}$ in the little group of $\Gamma$-Z, they can simply cross each other without opening a gap at the Fermi level, which forms a time-reversed pair of Dirac points at (0,0,$\pm k_D$) near the $\Gamma$ point. Interestingly, the Dirac points can be shifted along the $k_z$ axis by external strains that respect the $\mathcal{C}_{4z}$ symmetry. For better observation of the HOFAs, we increase the separation between two Dirac points $\Delta k=2k_D$ by applying a uniaxial tensile strain of 6\% along the $c$ axis, as shown in Fig.~\ref{fig:NaCuSe Structure}(c). Below, we utilize the strained system to illustrate the topological properties, unless otherwise specified.

To reveal the nontrivial higher-order topology in NaCuSe, we calculate the Wannier bands for $k_z$ slices on either side of the Dirac point. Due to the $\mathcal{C}_{4z}$ symmetry, it is sufficient to just consider $\nu_x(k_y)$. For generic slices ($k_z\neq 0,\pi$, or $\pm k_D$), four Wannier sectors are separated by Wannier gaps, as shown in Fig.~\ref{fig:NaCuSe Structure}(d,e). The higher-order topology is identified by the generalized WSP $p_y^{\nu_x^s}(k_z)$ parameterized as a function of $k_z$. As shown in Fig.~\ref{fig:NaCuSe Structure}(f), the gNBP component $\theta_y^{(4)}=2\pi p_y^{\nu_x^s}(k_z)$ {[$\overset{\mathcal{C}_{4z}}{=} 2\pi p_x^{\nu_y^s}(k_z)=\theta_x^{(4)}$]} has a nontrivial value of $\pi$ for the nonzero values of $k_z$ between two Dirac points ($0<|k_z|<k_D$). The sudden change of $\theta_y^{(4)}$ indicates that $k_z$ slices with hinge-localized HOFA states occur between the projection of bulk Dirac points, which agrees with our analysis of the change in filling anomaly between $k_z$ slices adjacent to the Dirac point \footnotemark[\value{footnote}]. The dramatic change of Wannier bands around the Dirac points can also be verified by directly scanning $\nu_x(k_z)$ adjacent to $k_z=\pm k_D$ in the $k_y=0$ plane \footnotemark[\value{footnote}]. It is noted that Wannier gaps vanish in the $\mathcal{T}$-invariant $k_z=0$ or $\pi$ slices, which makes the generalized WSP ill-defined. Instead, the $\mathbb{Z}_2$ topological invariants are well-defined in these planes and can identify the band topology \cite{huanghqSn,huanghqMgTa2N3,JiahengJAP}. It is found that $\mathbb{Z}_2=1$ for $k_z=0$ plane which exhibits the topology of a 2D TI, whereas $\mathbb{Z}_2=0$ for the $k_z=\pi$ plane. This identifies NaCuSe as a topologically nontrivial Dirac semimetal according to Ref.~\cite{NagaosaNC,PhysRevB.95.075135}.

\begin{figure}
  \centering  \includegraphics[width=1.0\linewidth]{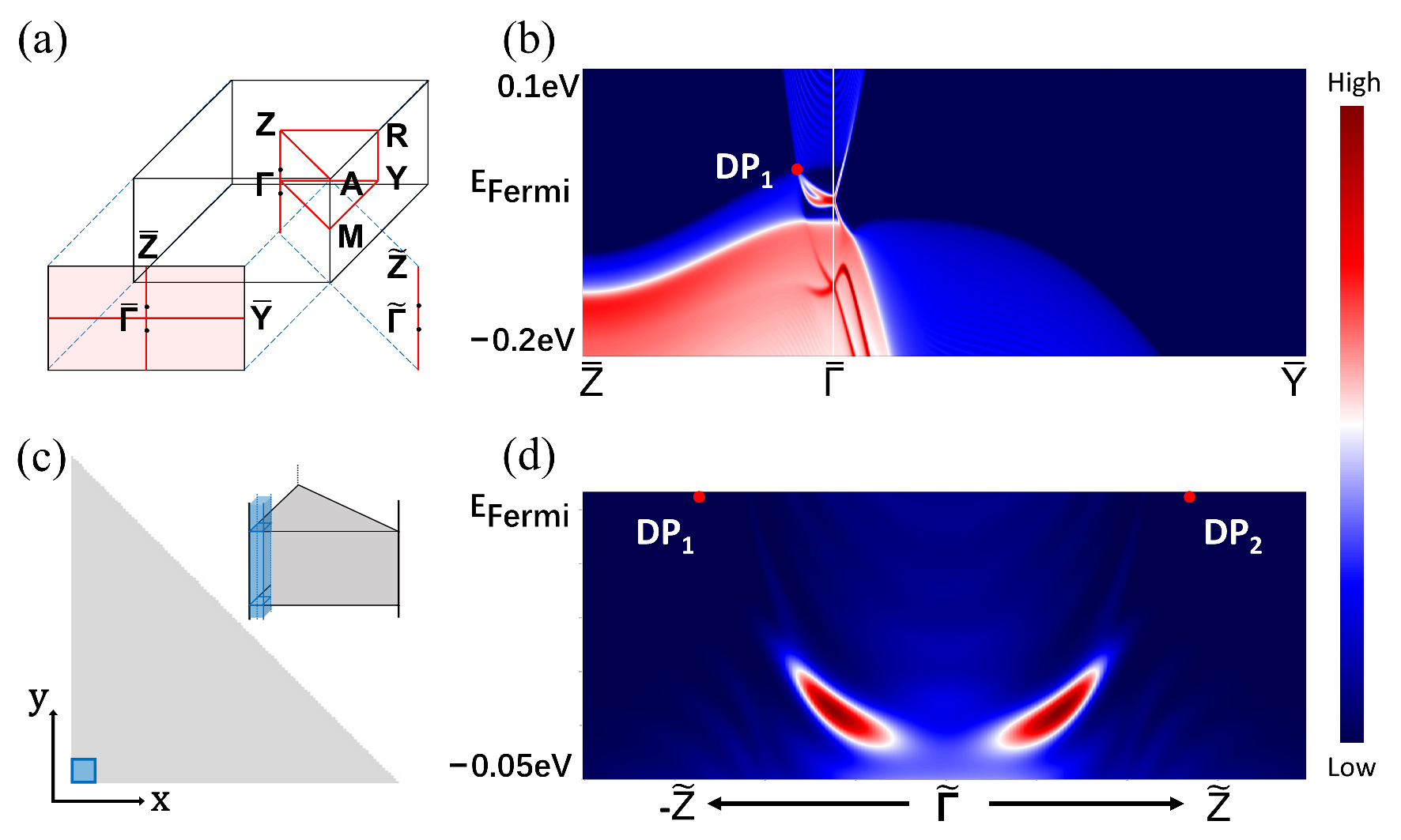}
  \caption{(a) Bulk, Surface, and Hinge Brillouin zones of NaCuSe. (b) (100) Surface spectrum of NaCuSe. (c) The right-angle prism-shaped nanorod for the calculation of the hinge spectrum. (d) Hinge spectrum of NaCuSe where HOFAs are clearly visible.}
  \label{fig:NaCuSe HOFA}
\end{figure}

Figure~\ref{fig:NaCuSe HOFA} shows our analysis of the surface and hinge spectra of the strained Dirac semimetal NaCuSe. As shown in Fig.~\ref{fig:NaCuSe HOFA}(a,b), there are some surface states on the (100) surface, which, however, are not necessitated to connect to surface projections of the bulk Dirac points \cite{Nie2022KagomeHOFA, Kargarian2015Are_FA_in_DSM_TP}. These states are not a topological consequence of the bulk Dirac points, but rather appear on the $\bar{\Gamma}$-$\bar{Y}$ line solely due to the nontrivial $\mathbb{Z}_2$ topology of the $k_z=0$ slice. In the spectrum of a single hinge in a right-angle prism-shaped nanorod [Fig.~\ref{fig:NaCuSe HOFA}(c,d)], we observe HOFA states in the vicinity of $k_z =0$ $(i.e., \tilde{\Gamma})$ that connect the hinge projections of the bulk Dirac points to the projections of 2D surface TI cones at $k_z=0$. This hinge mode is consistent with our analysis of the generalized WSP from the bulk electronic structure. Thus we confirm that the bulk-hinge correspondence in Dirac semimetals can be diagnosed by the change in generalized WSP.

In addition, we also reexamine the previously-studied Dirac semimetal KMgBi in space group 129. It has been identified that KMgBi is a Dirac semimetal with critically tilted Dirac cones \cite{le2017KMgBiCritical}, and its hinge spectrum exhibit HOFAs terminated at the hinge projection of the bulk Dirac points \cite{bernevig2020HOFA}. To determine the gNBP, we construct an \textit{ab initio} tight-binding model that respects the nonsymmorphic space group of KMgBi \footnotemark[\value{footnote}]. Surprisingly, our calculated conventional NBP $\theta^{(2)}(k_z)$ of $k_z$ slices are trivial on either side of Dirac points, which is different from previous results based on a simplified $k\cdot p$ Hamiltonian obtained by fitting to the first-principles band structures. Instead, we find that only the gNBP undergoes a change of $(\pi,\pi)$ when $k_z$ passes through the Dirac point, in agreement with the HOFAs terminating at the projected Dirac point. This again validates the necessity for the generalization of the nested Wilson loop formalism in Dirac semimetals with nonsymmorphic space groups.

\textit{Conclusion.}---We generalize the nested Wilson loop formalism to indicate the higher-order topology in 2D and 3D systems with nonsymmorphic symmetries. Due to symmetry constraints, Wannier bands are naturally split into multiple sectors and the gNBP of each sector can take quantized values, which serves as a new topological indicator to diagnose higher-order band topology. In particular, we find that the existence/absence of HOFAs in 3D Dirac semimetals is determined by the change in the gNBP of $k_z$ slices adjacent to Dirac points. Employing first-principles calculations, we demonstrate the HOFAs and gNBP in Dirac semimetals NaCuSe and KMgBi.

\begin{acknowledgments}
This work was supported by the National Key R\&D Program of China (No. 2021YFA1401600), the National Natural Science Foundation of China (Grant No. 12074006), and the top-notch student training in basic disciplines program 2.0 research project (Grant No. 20222005). H.Z. and W.D. acknowledge support from the Basic Science Center Project of NSFC (Grant No. 52388201) and the Beijing Advanced Innovation Center for Future Chip (ICFC). The computational resources were supported by the high-performance computing platform of Peking University.
\end{acknowledgments}

%\bibliography{type2dirac}
\providecommand{\noopsort}[1]{}\providecommand{\singleletter}[1]{#1}%

\end{document}